\documentclass{article}

\def\alf{\alpha}     
\def\gam{\gamma}   \def\dlt{\delta}
\def\veps{\varepsilon}   \def\vphi{\varphi}

\def\la{\langle}   \def\ra{\rangle}
   \def\dg{\dagger}

\def\pf{\mbox{\rm \small pf}}
\def\beq{\begin{equation}}
\def\eeq{\end{equation}}
\def\bea{\begin{eqnarray}}
\def\eea{\end{eqnarray}}
\def\nn{\nonumber}

\def\bt{\begin{tabular}}
\def\et{\end{tabular}}

\def\lb{\label}
\begin{document}

\title{\bf Nonlinear $n$-Pseudo Fermions}
\author{D. A. Trifonov\\
Institute for Nuclear Research and Nuclear Energetics,\\
 72 Tzarigradsko chaussee, Sofia, Bulgaria}
\maketitle
\begin{abstract}
Nonlinear pseudo-fermions of degree $n$ ($n$-pseudo-fermions) are introduced as (pseudo) particles with  creation and annihilation operators $a$ and $b$, $b \neq a^\dagger$, obeying the simple nonlinear anticommutation relation $ab+ b^n a^n =1$. The ($n+1$)-order nilpotency of these operators follows from the existence of unique (up to a bi-normalization factor) $a$-vacuum. Supposing  appropriate ($n+1$)-order  nilpotent  para-Grassmann variables and integration rules the sets of $n$-pseudo-fermion number states, and 'right' and 'left' ladder operator bi-overcomplete sets of coherent states are constructed. Explicit examples of  $n$-pseudo-fermion ladder operators are provided, and the relation of pseudo-fermions to finite-level pseudo-Hermitian systems is briefly considered.
\end{abstract}

\section{Introduction}

In the last decade or so a great deal of attention has been paid in literature to quantum systems with non-Hermitian (quasi-Hermitian, chrypto-Hermitian, $PT$-symmetric, pseudo-Hermitian) Hamiltonians  (see review articles \cite{Bender07, M-zadeh10} and references therein). More recently a considerable attention is paid to an alternative formalism for description of non-Hermitian systems, based on the concept of the so-called pseudo-bosons (PB) \cite{Trif09}--\cite{Calabrese} and pseudo-fermions (PF) \cite{M-zadeh04}--\cite{Maleki11}. Pseudo-bosons were originally introduced in \cite{Trif09}, where the first bi-overcomplete sets of  PB coherent states (CS) is constructed on the example of one-parameter family of PB. Mathematical refinement and further relevant examples of PB are due to Bagarello \cite{FB10a}--\cite{FB11a}. Para-Grassmann CS (nonnormalized  ladder-operator CS) for pseudo-Hermitian finite level Hamiltonian systems are constructed and discussed in \cite{Fasihi10, Maleki11}.

The notion of   pseudo-Hermitian fermion (phermion) was introduced by Mostafazadeh \cite{M-zadeh04}. Physical example of phermions is given in \cite{Cherbal'07}, where (phermions were called pseudo-fermions and) the first bi-overcomplete family of PF coherent states was established. The standard fermions and the pseudo-fermions so far considered \cite{M-zadeh04, Cherbal'07} are defined through linear in terms of the corresponding creation and annihilation operators anticommutation relations.

In the present paper we introduce {\it non-linear pseudo-fermions}\, of degree of nonlinearity $n$ ($n$-PF), $n$ being  positive integer. These are a non-Hermitian extension of the {\it non-linear fermions} ($n$-fermions) described in the previous paper \cite{Trif12}, and are relevant for description of finite level non-Hermitian quantum systems. In the next section we provide a brief summary of \cite{Trif12}. In the third section the  non-linear pseudo-fermions are introduced and several examples of are presented. In the fourth section 'left'and 'right' ladder operator CS are constructed. The $n$-PF  in  finite level pseudo-Hermitian systems are briefly considered in fifth  section.

\section{Nonlinear $n$-fermions}

The nonlinear $n$-fermions are defined \cite{Trif12} as particles  with annihilation and creation operators $A(n)$ and $A^\dg(n)$ satisfying the following nonlinear anticommutation relation \footnote{Such  relation  has been suggested (and realized for $n=2,3$) in ref.  \cite{Demichev} in the context of polynomial relations for the generators of the su$(2)$ Lie algebra.}
\begin{equation}\label{A alg}
A(n)A^\dg(n) + {A^\dg}^n(n)A^{n}(n) = 1,
\end{equation}
$n$ being a positive integer.
At $n=1$ the standard fermionic relations $aa^\dg + a^\dg a = 1$ are recovered, i.e.  $A(1)= a$. In this terminology the standard fermions are "$1$-fermions", or {\it linear fermions}.

Supposing the existence of a normalized vacuum state $|0\ra$ that is annihilated by $A(n)$, one can construct $n$ excited orthonormalized  states (Fock states),
\begin{equation}\label{|k>}
|k\ra = {A^\dg}^k(n)|0\ra,\quad k=0,1, \ldots, n,
\end{equation}
and deduce that $A(n)$ are nilpotent of order $n+1$: $A^{n+1}=0$.  The operators $A(n)$, $A^\dg(n)$ act on $|k\ra$ as raising and lowering operators with step $1$:
\begin{equation}\label{A|k>}
A(n)|k\ra =   |k-1\ra, \quad A^\dg(n) |k \ra = |k+1\ra.
\end{equation}
The corresponding number operator $N$, $N|k\ra =k|k\ra$, reads
\begin{equation}\label{N}
N(n) = A^\dg(n)A(n) + {A^\dg}^2(n)A^2(n) + \ldots + {A^\dg}^n(n)A^n(n),
\end{equation}
\begin{equation}\label{[A,N]}
[A(n),N(n)] = A(n), \quad [A^\dg(n),N(n)] = -A^\dg(n).
\end{equation}

In this way the state $|k\ra$, eq. (\ref{|k>}), can be regarded as a normalized state with $k$ number of $n$-fermions, $k=0,1,\ldots,n$.
There are no states with more than $n$ such particles.
So the degree of nonlinearity $n$ is the {\it order of statistics}  of our $n$-fermions.
The algebra spanned by  $A(n),A^\dg(n)$ and $N$, satisfying (\ref{A alg}) and (\ref{[A,N]}) could be called  {\it $n$-fermion algebra}.  At $n=1$ it coincides with the (standard) fermion algebra.
\medskip

{\bf Matrix realization.} One can check that the $n$-fermion algebra (\ref{A alg}) admits the following $(n+1)\times(n+1)$ matrix representation,
 \begin{equation}\label{A_{ik}}
 A(n)  = \left( \matrix{0&1&0&0&\ldots&0&0\cr 0&0&1&0&\ldots&0&0\cr
                       0&0&0&1&\ldots&0&0\cr
                      .&.&.&.&\ldots&.&.\cr 0&0&0&0&\ldots&0&1\cr
                       0&0&0&0&\ldots&0&0 } \right), \qquad
|0\ra = \left( \matrix{1&\cr 0\cr \cdot \cr \cdot \cr \cdot \cr 0 } \right) .
 \end{equation}
\medskip

In order to constructs eigenstates of $A(n)$ we adopt the following para-Grassmann algebra for the eigenvalues $\zeta$,
 \begin{equation} \label{zeta alg}
\zeta\zeta^* + \zeta^*\zeta = 0,\quad  \zeta^{n+1} = 0 = {\zeta^*}^{n+1},
\end{equation}
and the relations
 \begin{equation} \label{zeta A-comm}
 \zeta A(n) + A(n) \zeta = 0 = \zeta A^\dg(n) + A^\dg(n)\zeta, \quad \zeta |0\ra = |0\ra \zeta.
\end{equation}
These relations are most  simple and direct generalizations of  those for the standard fermion operators and their Grassmann eigenvalues \cite{Cahill}. They differ from para-Grassmann relations used e.g. in \cite{Fasihi10, Maleki11, Kibler}.  The para-Grassmann algebra (\ref{zeta alg}) admits $(n+1)\times(n+1)$ matrix representation \cite{Trif12}.

Using (\ref{zeta alg}) and (\ref{zeta A-comm}) the 'right' and 'left' (nonnormalized) eigenstates of $A(n)$ can be constructed \cite{Trif12},
\begin{eqnarray}\label{||zeta>_r}   
 A(n)||\zeta;n\ra_r &=& \zeta||\zeta;n\ra_r ,\nonumber\\
||\zeta;n\ra_r &=&  \sum_{k=0}^n (-1)^k \zeta^k |k\ra,
\end{eqnarray}
\begin{eqnarray}\label{||zeta>_l}    
A(n) ||\zeta;n\ra_l &=& ||\zeta;n\ra_l\zeta, \nn\\
||\zeta;n\ra_l &=& \sum_{k=0}^n (-1)^{[\frac{k+1}{2}]} \zeta^k |k\ra,
\end{eqnarray}
The normalized states are $|\zeta;n\ra_r=N_r||\zeta;n\ra_r$, $|\zeta;n\ra_l=N_l||\zeta;n\ra_l$ where $N_r = N_l = \sqrt{1-\zeta^*\zeta}$. However, in view of (\ref{zeta alg}), (\ref{zeta A-comm}),  the normalized states $|\zeta;n\ra_l$, unlike the 'right' $|\zeta;n\ra_r$,  cease being eigenstates of $A(n)$. This unsatisfactory property (and several other ones \cite{Trif12}) of $|\zeta;n\ra_l$  occurs for
other types of para-Grassmann 'left' eigenstates \cite{Fasihi10, Maleki11, Kibler, Cabra'06} as well.

The sets of both eigenstates $|\zeta;n\ra_r$ and $|\zeta;n\ra_l$ can resolve the identity operator,
\begin{equation}\label{res 1}
 \int d\zeta^* d\zeta\,  |\zeta;n\ra_r\,_r\la n;\zeta| = 1,
\end{equation}
\begin{equation}\label{res 2}
 \int d\zeta^* d\zeta\,  |\zeta;n\ra_l\,_l\la n;\zeta| = 1,
\end{equation}
if one adopt the following integration rules \cite{Trif12}
 \begin{equation}\label{intrul 1}
\int d\zeta^* d\zeta \, \zeta^i{\zeta^*}^k =  \delta_{ik} g_k(n),
\end{equation}
where  $g_k(n)$ are given by
\beq\lb{g_k}
g_k(n) = 1+\sum_{i=1}^{n-k}(-1)^{ki +\frac{i(i+1)}{2}} .
\eeq
For $ k=n, n-1, n-2, n-3$ we have
\beq\lb{g_n,g_(n-3)}
g_n =1,\quad g_{n-1} = 1 + (-1)^n,\quad  g_{n-2} = (-1)^{n-1},\quad  g_{n-3} = 0.
\eeq
Note the {\it different structure} of $g_k(n)$ for odd and even $n$ (i.e. for even and odd  dimension of the Hilbert space ${\cal H}_{n+1}$), the structure for odd $n$ being most simple. At $n=1$ the Berezin rules are reproduced \cite{Cahill}.  Thus the states $ |\zeta;n\ra_r$ and $ |\zeta;n\ra_l$ can be qualified as coherent states (CS)  -- the $n$-fermion ladder operator CS. At $n=1$ they reproduce the standard fermionic CS  \cite{Cahill}. The $n$-fermion  displacement-operator-like CS can also be constructed, this time the overcompleteness relation needing appropriate weight function $W(\zeta^*\zeta)$ \cite{Trif12}.

\section{Nonlinear pseudo-fermions}

The nonlinear pseudo-fermions of degree $n$ ($n$-PF) are defined as (pseudo) particles with non-Hermitian  annihilation and creation operators $a(n)$, $b(n)$, $b(n)\neq a^\dg(n)$, satisfying the $n$-nonlinear anticommutation relation
\begin{equation}\label{{a,b}}
a(n)b(n) + b^n(n) a^n(n) = 1.
\end{equation}
At $n=1$ and $b(1)=a^\#(1)$ the phermion relation  $aa^\# + a^\# a = 1$ is recovered \cite{M-zadeh04, Cherbal'07}.  In \cite{Trif09} a suggestion is made that if a pseudo-fermion operator $a(1)\equiv a$ admits a vacuum ($a|0\ra =0$) then $b^\dg$ also admits a vacuum and $b$ is $\eta$-pseudo adjoint to $a$, i.e. $b=a^\#$.  It appears that this suggestion could be made for $n>1$ as well.  In the above terminology the pseudo-fermions are "$1$-PFs" (or {\it linear} PFs).

Suppose that $|\psi_0\ra$ is annihilated by $a(n)$. Then we construct excited states $|\psi_k\ra$, $k=0,1,\ldots $
\beq\lb{psi_k}
|\psi_k\ra = b^k(n)|\psi_0\ra,
\eeq
on which $b$ and $a$ act as raising and lowering operators with step $1$,
\beq\lb{b a psi_k}
b(n)|\psi_k\ra = |\psi_{k+1}\ra,\qquad a(n)|\psi_k\ra = |\psi_{k-1}\ra.
\eeq
The process is terminated at $k=n$, i.e. $b^{n+1}|\psi_0\ra =0$, and this follows from the anticommutation relation (\ref{{a,b}}). Indeed,
take $|\psi_{n+1}\ra := b^{n+1}|\psi_0\ra$, and multiply it by  $a(n)b(n) + b^n(n) a^n(n)$. Using (\ref{{a,b}})--(\ref{b a psi_k}) we obtain
$$|\psi_{n+1}\ra =  (a(n)b(n) + b^n(n) a^n(n) |\psi_{n+1}\ra = a(n)b(n)|\psi_{n+1}\ra + b^n(n) a^n(n)) |\psi_{n+1}\ra =  2|\psi_{n+1}\ra,$$
which is possible iff $ |\psi_{n+1}\ra := b^{n+1}|\psi_0\ra = 0$. This means that in the space ${\cal H}_{n+1}$ spanned by the $n+1$ vectors $|\psi_k\ra$ the operators $a$ and $b$ are nilpotent (matrices) of order $n+1$, i.e. $b^{n+1}=a^{n+1}=0$. So the set of $|\psi_k\ra$ is a basis in ${\cal H}_{n+1}$, but since $b\neq a^\dg$ this basis  is not orthogonal.

One can check (using the anticommutation relation (\ref{{a,b}})) that the states $|\psi_k\ra$ are eigenstates of the non-Hermitian operator
\beq\lb{Npf}
N_{\rm pf}(n) = b(n)a(n) + b^2(n)a^2(n) + \ldots + b^n(n)a^n(n)
\eeq
with eigenvalue $k$: $N_{\rm pf}(n)|\psi_k\ra = k|\psi_k\ra$.  So $N_{\rm pf}(n)$ plays the role of (non-Hermitian) number operator for $n$-pseudo-fermions. One can verify that
\beq\lb{alg1}
[a(n),N_{\rm pf}(n)] = a(n),\quad [b(n),N_{\rm pf}(n)] = -b(n).
\eeq

Since all the $n+1$ distinct eigenvalues of $N_{\rm pf}(n)$ are real (nonnegative integers $k$) the eigenvalues of its Hermitian conjugate $N_{\rm pf}^\dg(n)$ are the same (real nonnegative integers $k$). Denoting the corresponding eigenvectors as $|\vphi_k\ra$ we write
\beq\lb{vphi_k}     
N_{\rm pf}^\dg |\vphi_k\ra = k |\vphi_k\ra.
\eeq
The nonorthogonal eigenvectors $|\vphi_k\ra$ can be similarly constructed from the $N_{\rm pf}^\dg(n)$-lower eigenstate $|\vphi_0\ra$ by means of the raising operator $a^\dg$  (using $b^\dg a^\dg + {a^\dg}^n {b^\dg}^n=1$ and $[a^\dg(n),N_{\rm pf}^\dg(n)] = -a^\dg(n),\quad [b^\dg(n),N_{\rm pf}^\dg(n)] = b^\dg(n)$),
\beq\lb{vphi_k}
|\vphi_k\ra = {a^\dg}^k(n)|\vphi_0\ra.
\eeq
It is worth noting at this place that the existence of the nontrivial solution $|\vphi_0\ra$ of the equation $N_{\rm pf}^\dg(n)|\vphi_0\ra = 0$ follows from the well known property of systems of (here $n+1$) linear homogeneous algebraic equations $A\vec{x} =0$: the nontrivial solution $\vec{x}$ exists iff det$A=0$. Indeed, if $N_{\rm pf}(n)|\psi_0\ra = 0$, $|\psi_0\ra \neq 0$, then the matrix of $N_{\rm pf}(n)$ has a vanishing determinant, and so is the case with the determinant of the matrix $N_{\rm pf}^\dg(n)$. Therefore the equation $N_{\rm pf}^\dg(n)|\vphi_0\ra = 0$ admits nontrivial solution.

Moreover, the vector $|\vphi_0\ra$ should be annihilated by the operator $b^\dg$: $b^\dg |\vphi_0\ra = 0$. This can be easily proven as follows. Applying $N_{\rm pf}^\dg$ to $b^\dg|\vphi_0\ra$ and using  $N_{\rm pf}^\dg |\vphi_k\ra = 0$ and $[b^\dg(n),N_{\rm pf}^\dg(n)] = b^\dg(n)$ we find that $b^\dg|\vphi_0\ra$ is an eigenstate of $N_{\rm pf}^\dg$ with the new eigenvalue $-1$. Therefore $b^\dg|\vphi_0\ra$ should be orthogonal to all eigenstates $|\psi_k\ra$ of $N_{\rm pf}$. The set $\{|\psi_k\ra\}$ form a basis in  ${\cal H}_{n+1}$, therefore $b^\dg|\vphi_0\ra=0$.  Thus if $a(n)$-vacuum $|\psi_0\ra$ exists, then $b^\dg(n)$-vacuum also exists (and vice versa).

The orthogonality of eigenstates $|\psi_i\ra$ and $|\vphi_j\ra$ of non-Hermitian operators $H$ and $H^\dg$ with different real eigenvalues $\veps_i$, $\veps_j$, used in the above, is a known fact, but for the sake of completeness let us provide here its short proof: $\la \vphi_j|H^2\psi_i\ra = \la \vphi_j|\psi_i\ra \veps_i^2 = \la \vphi_j|\psi_i\ra \veps_i\veps_j$. If $\veps_i\neq \veps_j$ the last equality is possible iff $\la \vphi_j|\psi_i\ra=0$. The states $|\psi_k\ra$ and $|\vphi_k\ra$ corresponding to equal eigenvalues can be bi-normalized, so that we have a bi-orthonormalized system of $n$-pseudo-fermion states,
\beq\lb{bi-norm}
\la \vphi_k|\psi_j\ra = \delta_{kj}.
\eeq
Herefrom it follows that $|\psi_j\ra = \eta|\vphi_j\ra$, where $\eta = \sum_k |\vphi_k\ra\la\vphi_k|$, the inverse operator being $\eta^{-1} =
\sum_k |\psi_k\ra\la\psi_k|$.  Next, one can readily verify (using (\ref{bi-norm}) and the basis $\{|\psi_k\ra\}$), that the sum $\sum_k|\psi_k\ra\la\vphi_k|$ acts on any state $|\psi\ra$ as unit operator,
\beq\lb{1}   
1= \sum_k|\psi_k\ra\la\vphi_k|.
\eeq
Finally we note that the $n$-PF creation operator $b(n)$ is $\eta$-pseudo-adjoint to $a(n)$: $b(n)=\eta^{-1}a^\dg(n)\eta =: b^\#$. This can be verified by applying  $\eta^{-1}a^\dg(n)\eta$ to  $|\psi_k\ra$ (the basis vectors in ${\cal H}_{n+1}$) and see that these actions are the same as those of $b(n)$.
\bigskip

{\bf Three examples} of $n$-PFs:
\medskip

{\bf (1)} $n=2$.
\bea
a &=& \alf A^\dg(2) + \beta {A^\dg}^2(2) A(2), \nn\\
b &=& \frac{1}{\alf+\beta} A(2) + \frac{\beta}{\alf(\alf+\beta)} A^2(2) A^\dg(2),\nn
\eea
where $A(2),\, A^\dg(2)$ are Hermitian ladder operators of $2$-fermions (see section 2). One can check the validity of  all the required relations for $2$-PF with any $\beta$ and  nonvanishing $\alf$, $\alf \neq -\beta$:
\bea
b  &\neq& a^\dg, \nn \\
ab+b^2a^2&=&1, \quad a^3=0=b^3, \nn\\
a|\psi_0\ra &=& 0 \longrightarrow  \la \psi_0| = (0,0,p^*), \nn\\
b^\dg |\vphi_0\ra &=& 0 \longrightarrow \la\vphi_0| = (0,0,1/p), \nn
\eea
$p$ being any nonvanishing complex number.
\medskip

{\bf (2)} $n=3$. Any nonvanishing $\alf$, $\beta$, $\gam$, $\delta$, $p$\,:
\bea
a &=& \left(\matrix{0& \alf& 0& 0\cr 0& 0& 0& 0\cr \dlt \gam& 0 &\beta/\dlt& -\beta/\dlt^2\cr 
0& 0& \beta& -\beta/\dlt }\right), \quad
b = \left(\matrix{0& 0& 1/\gam\dlt & -1/\gam\dlt^2 \cr 1/\alf& 0& 0& 0\cr 0& 0& 0& 1/\beta\cr 0&0 & 0 &0}\right), \nn\\[2mm]
\quad\la \psi_0| &=& (0, 0, p^*, p^*\dlt^*),\qquad\quad \la \vphi_0| = (0, 0,0, 1/p\dlt). \nn
\eea
\medskip

{\bf (3)} $n > 1$.  Any nonvanishing $\alf_i$, $i=1,2, \ldots, n$,  $p$\,:
\bea
a &=& \left(\matrix{0& \alf_1& 0& \ldots& 0\cr 0& 0& \alf_2& \ldots&0\cr \cdot& \cdot& \cdot& \cdot&\cdot\cr
0& 0& \cdot& \ldots& \alf_n\cr 0& 0& \cdot&\ldots& 0}\right), \quad
b = \left(\matrix{0& 0& 0& \ldots& 0& 0\cr \alf_1^{-1}& 0& 0& \ldots& 0&0\cr 0& 0& \alf_2^{-1}& \ldots& 0& 0\cr
\cdot& \cdot& \cdot& \ldots& \cdot& \cdot&\cr 0& 0& 0&\ldots& \alf_n^{-1} &0}\right),  \nn\\[2mm]
\quad\la \psi_0| &=& (p^*, 0, \dots, 0),\qquad\qquad\qquad \la \vphi_0| = (1/p, 0, \ldots, 0). \nn
\eea
\medskip

\section{$n$-PF ladder  operator eigenstates}

The two $n$-pseudo-fermion lowering and raising ladder operators $a(n)$ and $b(n)$ can be diagonalized using the same para-Grassmann variables $\zeta$ as for the $n$-fermions,  eq. (\ref{zeta alg}), and the following relations:
\bea
\{a(n),\zeta\} &=& 0 = \{a(n),\zeta^*\}, \quad \{b(n),\zeta\} = 0 = \{b(n),\zeta^*\},\\
\{\zeta,|\psi_0\ra\} &=& 0 = \{\zeta^*,|\psi_0\ra\}, \quad \{\zeta,|\psi_0\ra\} = 0 = \{\zeta^*,|\psi_0\ra\}.
\eea
 For the non-bi-normalized 'right' and 'left' eigenstate we find
\bea\lb{pf_r}
a(n)||\zeta;n,\pf\ra_r &=& ||\zeta;n,\pf\ra_r \zeta, \nn\\
||\zeta;n,\pf\ra_r &=& \sum_{k=0} (-1)^k \zeta^k|\psi_k\ra,
\eea
\bea\lb{pf'_r}
b^\dg(n)||\zeta;n,\pf \ra_r' &=& ||\zeta;n, \pf \ra_r' \zeta, \nn\\
||\zeta;n,\pf \ra_r' &=&  \sum_{k=0} (-1)^k \zeta^k|\vphi_k\ra,
\eea

\bea\lb{pf_l}    
a(n)||\zeta;n,\pf\ra_l &=& \zeta||\zeta;n,\pf\ra_l, \nn\\
||\zeta;n,\pf\ra_l &=& \sum_{k=0} (-1)^{[\frac{k+1}{2}]} \zeta^k|\psi_k\ra,
\eea
\bea\lb{pf'_l}    
b^\dg(n)||\zeta;n,\pf \ra_l' &=& \zeta||\zeta;n,\pf \ra_l', \nn\\
||\zeta;n,\pf \ra_l' &=&  \sum_{k=0} (-1)^{[\frac{k+1}{2}]} \zeta^k|\vphi_k\ra.
\eea
The bi-normalized states satisfying the relations
\beq
\,_i\la \pf,n;\zeta|\zeta;n, \pf\ra'_i = 1,\quad i = l, r,
\eeq
are $|\zeta;n, \pf\ra_i ={\cal N}_i|\zeta;n, \pf\ra_i$, $|\zeta;n, \pf\ra_i' ={\cal N}_i'|\zeta;n, \pf\ra_i'$, where ${\cal N}_i{\cal N}_i' = \sqrt{1-\zeta^*\zeta}$. If $b=a^\dg$ they reproduce the normalized $n$-fermion eigenstates. Note however that the 'left' eigenstates $||\zeta;n, \pf\ra_l$, $||\zeta;n, \pf\ra_l'$ when bi-normalized cease being eigenstates of $a$ and $b^\dg$.  This feature is typical for all 'left' parafermionic eigenstates.

Next we look for bi-overcompleteness relations.  It turned out that the families of both 'left' and 'right' $n$-pseudo-fermion eigenstates form the bi-overcomplete sets of states with respect to the same new integration relations (\ref{intrul 1}), (\ref{g_k}). Instead of (\ref{res 1}), (\ref{res 2}) we now have the bi-overcompleteness relations
\beq\lb{res 3}
 \int d\zeta^* d\zeta\,  |\zeta;n,\pf \ra_r' \,_r\la \pf,n;\zeta| = 1,
\end{equation}
\begin{equation}\label{res 4}
 \int d\zeta^* d\zeta\,  |\zeta;n,\pf \ra_l' \,_l\la {\rm pf}, n;\zeta| = 1.
\end{equation}
In view of the above bi-overcompleteness relations the two sets  $\{|\zeta;n,\pf\ra_r, \,\, |\zeta;n,\pf\ra_r'\}$ and  $\{|\zeta;n,\pf\ra_l, \,\, |\zeta;n,\pf\ra_l'\}$ can be qualified as $n$-pseudo-fermion 'left' and 'right' ladder operator coherent states (CS). At $n=1$ they recover the bi-overcomplete sets of pseudo-fermionic CS  \cite{Cherbal'07} ($1$-pseudo-fermion, or linear pseudo-fermion CS in the present terminology).

\section{$n$-PF and  non-Hermitian systems}

In this section we briefly consider the relations of $n$-pseudo-fermions with finite level (non-Hermitian) quantum systems. For Hermitian systems similar relations are discussed in \cite{Trif12}.

Consider a system with finite number of non-degenerate (possibly not equidistant) 'energy'  levels ($n+1)$ levels) $\veps_k$, $k=0,1,\ldots,n$, and let $|\psi_k\ra$ be the corresponding wave functions. Denote the finite dimensional Hilbert space spanned by $|\psi_k\ra$ as ${\cal H}_{n+1}$.  If $|\psi_k\ra$ are not orthogonal to each other they could be regarded as eigenstates of some non-Hermitian Hamiltonian $H$: $H|\psi_k\ra = \veps_k|\psi_k\ra$. In a more general setting the eigenvalues $\veps_k$ may be complex quantities. If $\veps_k$ are real or come in complex conjugate pairs then $H$ is pseudo-Hermitian \cite{M-zadeh10}: $H= \eta^{-1}H^\dg \eta =:H^\#$, where $\eta$ is some Hermitian operator. For $H^\dg$ the eigenvalue relations are $H^\dg |\vphi_k\ra = \veps_k^*|\vphi_k\ra$. For different $k,\, k'$ the states $|\psi_k\ra$ and $|\vphi_{k'}\ra$ are orthogonal and for equal $k$ they can be bi-normalized, $\la \psi_k|\vphi_{k'}\ra = \delta_{kk'}$.

The general lowering and raising operators between levels are of the form \cite{Fasihi10}
\beq
a(n;\vec{\rho}) = \sum_{k=0}^{n-1} \sqrt{\rho_k}|\psi_k\ra\la \vphi_{k+1}| , \qquad
b(n;\vec{\rho})  = \sum_{k=0}^{n-1} \sqrt{\rho_k}\, |\psi_{k+1}\ra\la \vphi_k|.
\eeq
where $\rho_k$ are arbitrary complex (dimensionless) quantities.  Such operators with $\rho_k = [[k+1]] = (q^{k+1}-q^{-k-1})/(q-1/q)\equiv \rho^{(q)}_k$, $q=\exp(i\pi/(n+1))$  were considered in \cite{Maleki11}.  At $\rho_k=\veps_{k+1}$ we find that  $b(n;\vec{\veps})a(n;\vec{\veps})|\psi_k\ra = \veps_k |\psi_k\ra$, which means that these operators factorize the Hamiltonian,
\beq
H =  b(n;\vec{\veps})a(n;\vec{\veps}).
\eeq
If $\veps_0 \neq 0$ then $H = b(n;\vec{\veps})a(n;\vec{\veps}) + \veps_0$. Note that here $H$ is dimensionless, along with $\veps_k$ and operators $a(n)$ and $b(n)$.  

At $\rho_k=1$ the operators  $a(n;1)$, $b(n;1)$ obey the relation (\ref{{a,b}}), i.e. $a(n;1)$, $b(n;1)$ are $n$-pseudo-fermion ladder operators for the ($n+1$)-level quantum system: $a(n;1) = a(n)$, $b(n;1)=b(n)$. The general ladder operators  $a(n;\vec{\rho})$, $b(n;\vec{\rho})$ can be expressed in terms of $a(n)$, $b(n)$ as
 \beq\lb{arho- a}
a(n;\vec{\rho}) = \sigma_0 a(n) +(\sigma_1-\sigma_0)b(n) a^2(n) +\ldots + (\sigma_{n-1}-\sigma_{n-2}) b^{n-1}(n)a^n(n),
\eeq
\beq\lb{brho- b}
b(n;\vec{\rho}) = \sigma_0 b(n) + (\sigma_1- \sigma_0)b^2(n)a(n) +\ldots + (\sigma_{n-1} - \sigma_{n-2})b^n(n)a^{n-1}(n).
\eeq
where, to shorten the equations we have put $\sqrt{\rho_k} \equiv \sigma_k$. 
 
It is not difficult to check that $H$ commutes with the $n$-PF number operator $N_{\rm pf}(n) = b(n)a(n) + \ldots + b^n(n)a^n(n)$. In view also of the fact that $|\psi_k\ra$ are eigenstates of $N_{\rm pf}(n)$ with eigenvalues $k$ one could interpret the energy value $\veps_k$ as a sum of the energies $\veps_k/k$ of $k$ number of $n$-pseudo-fermions.  If the spectrum of $H$ is equidistant, then $H$ is proportional to $N_{\rm pf}(n)$.

 Following the scheme developed in section 3 (and in \cite{Trif12} for Hermitian case) one can construct bi-overcomplete 'left' and 'right'  ladder operator eigenstates for $a(n;\vec{\rho})$ and $b^\dg(n;\vec{\rho})$ using  the para-Grassmann algebra (\ref{zeta alg}) and the integration rules
(\ref{intrul 1}).  Bi-overcomplete sets of para-Grassmann 'left'  (nonnormalized) eigenstates of $a(n;\vec{\rho})$ and $b^\dg(n;\vec{\rho})$ were constructed in \cite{Fasihi10} using different paragrassmannian variables and integration rules. In \cite{Maleki11} overcomplete families of eigenstates of $a'= q^{N/2}a(n;\vec{\rho}^{(q)})$ (where, in our notations,  $[[N]]= b(n;\vec{\rho}^{(q)}) a(n;\vec{\rho}^{(q)})$) and $c'=  q^{N'/2}b^\dg(n;\vec{\rho}^{(q)})$ (where  $[[N']] = a^\dg(n;\vec{\rho}^{(q)}) \, b^\dg(n;\vec{\rho}^{(q)}$) are built up using also different paragrassmannian variables and integration rules.

\medskip
\section*{Conclusion}
We have introduced nonlinear pseudo-fermions of degree $n$ ($n$-pseudo-fermions) as (pseudo) particles with non-Hermitian  annihilation and creation operators $a(n)$, $b(n)$, $b(n)\neq a^\dg(n)$, satisfying the $n$-nonlinear anticommutation relation $a(n)b(n) + b^n(n) a^n(n) = 1$. A pair of non-Hermitian operators $a$, $b\neq a^\dg$ could represent $n$-pseudo-fermion if they obey the relation $ab + b^n a^n = 1$ and $a$ admits a nontrivial ground state $|\psi_0\ra$, $a|\psi_0\ra =0$. Then $b^\dg$-vacuum also exists, $a$ and $b$ are nilpotent of order $n+1$, and bi-orthonormalized set of Fock states and bi-overcomplete sets of 'left' and 'right' CS can be constructed, as we have done this in the paper, using appropriately defined new paragrassmannian variables and integration rules.  At $b=a^\dg$ (the Hermitian case) $n$-pseudo-fermion CS recover  the $n$-fermion CS \cite{Trif12}, and  at $n=1$ both 'left' and 'right' $n$-pseudo-fermion CS reproduce the pseudo-fermionic CS of ref. \cite{Cherbal'07}. Different kind of (nonnormalized)  para-Grassmann 'left' CS for finite level pseudo-Hermitian systems are considered in \cite{Fasihi10, Maleki11}.

Three different families of $n$-pseudo-fermion operators have been provided as examples for $n=2$, $n=3$ and any $n>1$. The $n$-pseudo-fermion operators can be introduced for any pseudo-Hermitian system with non-degenerate finite number of energy levels $\veps_k$. The $n$-pseudo-fermion number operator commutes with the corresponding non-Hermitian Hamiltonian and thereby the energy $\veps_k$ could be regarded as  a sum of energies of $k$ number of pseudo-particle (or pseudo-excitations). If the energy levels are equidistant then the Hamiltonian is proportional to the pseudo-fermion number operator.

\end{document}